\documentclass[11pt,a4paper]{article}
\usepackage[margin=23mm]{geometry}
\usepackage[T1]{fontenc}
\usepackage{lmodern,microtype,amsmath,booktabs,tabularx,array,graphicx,float,placeins,fancyhdr,xcolor}
\usepackage[colorlinks=true,linkcolor=blue!45!black,citecolor=blue!45!black,urlcolor=blue!45!black]{hyperref}
\usepackage{enumitem}
\newcolumntype{Y}{>{\raggedright\arraybackslash}X}
\hypersetup{pdftitle={CALLIOPE: A Source-Grounded Oral Assessment System and Synthetic Readiness Evaluation},pdfauthor={Nizam Kadir}}
\title{\textbf{CALLIOPE:}\\A Source-Grounded Oral Assessment System\\and Synthetic Readiness Evaluation}
\author{Nizam Kadir\thanks{Provisional author line for review; final authorship and declarations require author confirmation.}\\\small Science, Mathematics and Technology\\\small Singapore University of Technology and Design, Singapore\\\small\texttt{nizam\_kadir@mymail.sutd.edu.sg}}
\date{25 September 2026}
\begin{document}
\maketitle
\begin{abstract}
Oral assessment with generative AI requires more than a conversational interface: educators must connect a spoken response to its source material, scoring criteria, model outputs and subsequent human judgement. This technical report presents CALLIOPE, a source-grounded oral assessment system integrating versioned instructional material, learner-turn recording and transcription, adaptive questioning, two-provider rubric scoring, educator review and exportable evidence. We examine implementation and retained synthetic verification records from 25 September 2026. Three spoken fixtures and a silence control were exercised across two release runs. In the final release rehearsal, the spoken fixtures received aggregate AI scores of 100, 62 and 8 out of 100; two elicited provider-disagreement flags. All eight retrieved audio files across the two runs were byte-identical to their inputs. These observations establish operation of specific exercised paths, not scoring validity or learning gains. A later zero-traffic candidate added version-bound consent checks, insert-only first-pass rating receipts and separately authorised coded exports, supported by local regression tests but not a new full live research workflow evaluation. We distinguish deployed functionality, candidate safeguards and remaining recovery, concurrency and study-operation requirements. The contribution is an inspectable response-to-review workflow and a release-specific account of what its engineering evidence does, and does not, establish. No human-participant outcomes are reported. OpenAI Codex assisted with technical verification, evidence synthesis and manuscript preparation.
\end{abstract}

\section{Introduction}
An oral explanation can expose how a learner connects concepts, justifies a claim and responds to a probing question. Automating parts of that conversation creates a second problem: an educator must be able to inspect the evidence behind the resulting feedback or score. A fluent response from an assessment agent is not itself evidence that the learner was assessed accurately. A successfully stored recording is likewise not evidence that the assessment improves learning.

CALLIOPE addresses the engineering continuity between assessment configuration, spoken evidence, machine evaluation and educator review. An educator supplies source material, concepts and rubric descriptors. A learner responds through an oral-session interface. The system retains configured learner-turn audio, transcribes responses, retrieves relevant source passages, generates follow-up questions and produces structured rubric scores. Review and export workflows expose different parts of this record to authorised roles.

This report examines the implementation and synthetic verification record available on 25 September 2026. It is a systems and readiness evaluation, not an educational intervention trial. The four unique fixtures are three machine-generated spoken responses and a silent input. Repeating them across two release runs does not create six independent learners, and a planned study of approximately 50 students is not an observed sample in this report.

The contribution has three parts: (1) an integrated, source-versioned response-to-review workflow; (2) an explicit separation of machine scoring, human rating and research export; and (3) a release-specific evidence account linking reported observations to their testing scope. We do not propose a new foundation model, claim the first AI oral examiner, or establish instructor-equivalent grading. The report's central question is how much readiness can be supported by the available evidence before a participant study, and which inferences remain unsupported.

\section{Assessment Foundations and Related Work}
Formative feedback is useful when it helps a learner understand goals, assess current work and decide what to do next, rather than merely receiving a numerical judgement \cite{nicol}. Accordingly, CALLIOPE's formative workflow combines explanations and probing questions with an educator-defined rubric. Whether learners subsequently improve requires an independent outcome measure; the interface alone cannot establish this benefit.

Oral-assessment interpretation also depends on task design and the relationship between the observed performance and the intended construct \cite{joughin}. Here, content accuracy, explanatory depth and reasoning concern expressed disciplinary understanding. The labels ``communication clarity'' and ``metacognitive awareness'' must not be treated as validated psychological measurements simply because a model returns values for them. A content-based rubric is distinct from pronunciation scoring or inference about identity, emotion or personality.

Retrieval-augmented generation provides a general rationale for supplying external evidence to a language model \cite{lewis}. CALLIOPE's implementation uses versioned source chunks and lexical retrieval; it is not an implementation or evaluation of the dense-retrieval model in that foundational work. Retrieval exposes which source passages were supplied, but neither a retrieved passage identifier nor a plausible generated rationale establishes semantic correctness.

Recent voice-AI assessment work already demonstrates related system patterns. Ipeirotis and Rizakos describe Viva, including voice examinations and multi-model grading \cite{viva}. Their operational evidence does not remove the need for independent assessment validation in a different system and course. CALLIOPE's two-provider procedure is independent scoring followed by arithmetic aggregation and disagreement flagging, not a multi-round debate protocol. Khan, Liow and Ang examine AI versus teaching-assistant assessment in design-thinking poster work \cite{khan}; that exploratory visual-assessment context is neither an oral CALLIOPE evaluation nor evidence of CALLIOPE learning gains.

LLM judges can exhibit position, verbosity and other biases \cite{zheng}. Multiple models may share errors, and averaging their scores cannot establish validity. Speech recognition adds another potential source of unequal error before scoring begins; published disparities in commercial ASR motivate evaluation across actual speakers and conditions \cite{koenecke}. These literatures motivate evidence preservation and human review, not a claim that either concern is resolved by the present implementation.

\section{System Design and Implementation}
\subsection{Assessment configuration and source grounding}
CALLIOPE is a browser-based application with server-side assessment, transcription, dialogue, evaluation and review routes. It separates learner, educator and administrative responsibilities. The educator configures a question, time allowance, formative or consequential use, rubric weights and descriptors, required concepts, terminology and source material. Published assessment sources are versioned so that subsequent changes to teaching materials need not silently change the evidential basis of an existing assessment.

The retrieval implementation ranks source chunks using lexical term overlap and frequency. The evaluated responses retain retrieved chunk identifiers, making it possible to connect an output with the source version supplied to it. Citation checks use text matching and term-overlap heuristics. These are provenance and consistency mechanisms, not a semantic entailment test. A retrieved source can be incomplete, inappropriate or itself incorrect; a response can cite a relevant passage while drawing an unsupported conclusion.

Table~\ref{tab:workflow} summarises the workflow and the boundaries of its evidence. The educator retains responsibility for the adequacy of source material and rubric descriptors. The biological cell-comparison material used in this report is a technical fixture, not a faculty-approved course assessment.

\begin{table}[htbp]\centering\small
\caption{Response-to-review workflow. A retained artefact is not automatically a validated measurement.}\label{tab:workflow}
\begin{tabularx}{\textwidth}{p{27mm}YY}\toprule
Stage & Implemented record or operation & Interpretation boundary\\\midrule
Configuration & Question, versioned source chunks, rubric, assessment mode & Requires disciplinary review and aligned task design.\\
Speech capture & Learner-turn audio and transcription & Not continuous room audio or a recording of both speakers.\\
Dialogue & Retrieved passages and generated follow-up question & Source access does not guarantee factual or pedagogical correctness.\\
AI evaluation & Six dimensions, provider status, disagreement and aggregate score & Model output is not an independent reference grade.\\
Educator review & Review state, rating inputs and publication workflow & Mechanical publication tests do not constitute human rating.\\
Research export & Candidate coded, explicit-scope package with exclusions & Requires study-specific authorisation and operational validation.\\\bottomrule
\end{tabularx}
\end{table}

\subsection{Oral interaction, transcription and evidence retention}
\begin{figure}[htbp]\centering
\includegraphics[width=\textwidth]{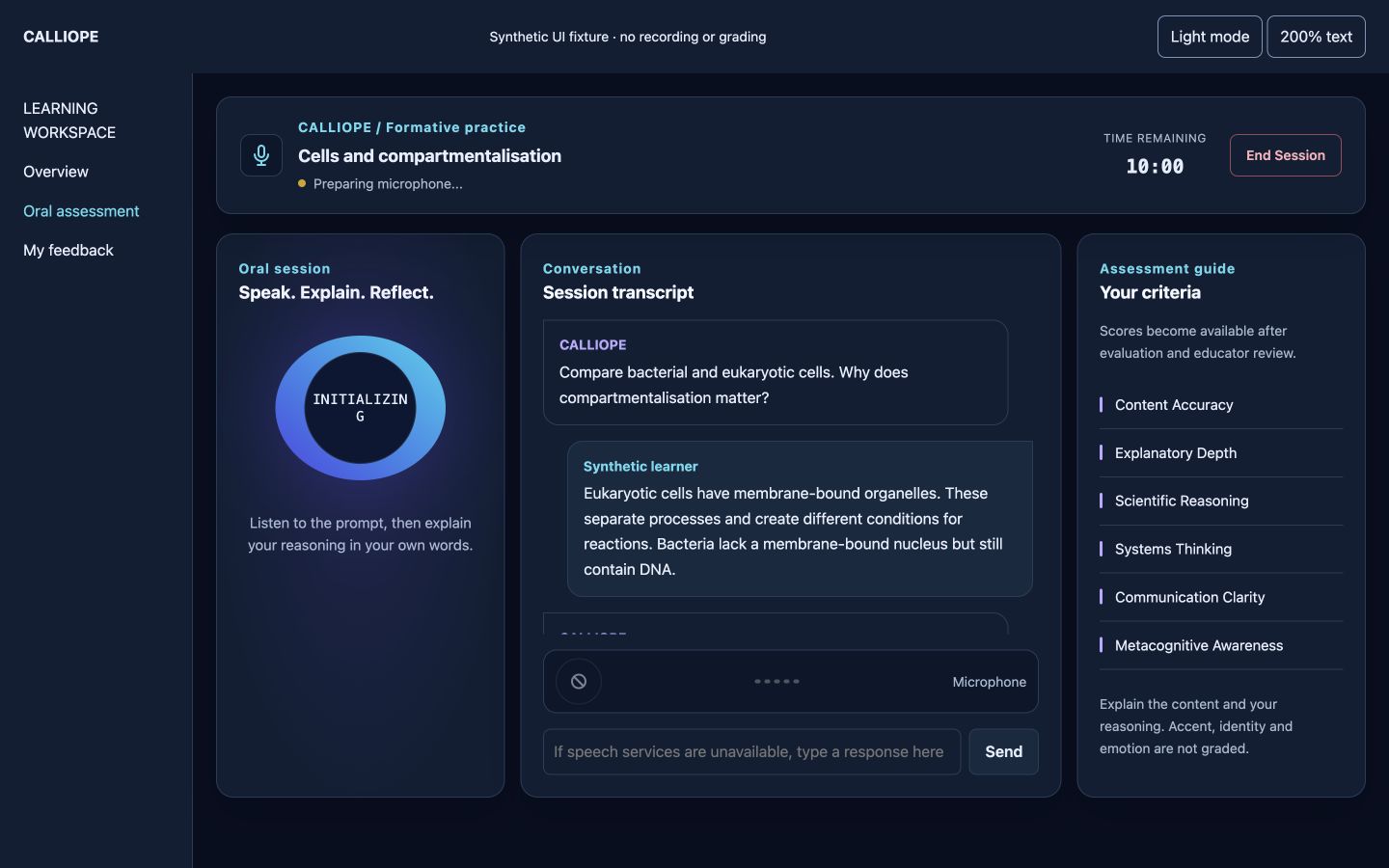}
\caption{Archived CALLIOPE desktop interface with synthetic cell-comparison text. The banner identifies a visual fixture with no recording or grading. This screenshot illustrates layout and criteria visibility, not an observed learner session or accessibility certification.}\label{fig:ui}
\end{figure}

The learner interface presents the assessment question, conversation transcript, session controls and assessment criteria. It supports light and dark themes and responsive layout. Figure~\ref{fig:ui} shows an archived desktop visual fixture. Its simulated transcript and microphone-initialisation display are interface evidence only; they are not evidence of live microphone operation.

The recorded pipeline separates audio, recognised text and downstream evaluation. In the retained rehearsal, speech-to-text responses identify \texttt{whisper-1}. The system also exposes observable speech descriptors and an ASR confidence field. The scoring prompt instructs evaluators not to use accent, pauses, pitch, loudness, emotion, identity or other such attributes in rubric scores. This instruction is a design constraint, not a demonstrated fairness result. None of the retained ASR responses contains a measured reference word-error rate; confidence values must not be reported as transcription accuracy.

Audio replay is permission controlled and represents retained learner-turn clips. Checkpointing permits restoration of already saved session state. However, recovery of saved text and replay of uploaded audio are different from durable recovery of an unsent recording. The latter, together with shared-device cleanup and interruption handling, remains a release-readiness requirement discussed in Section~\ref{sec:limits}.

\subsection{Two-provider rubric scoring}
The implementation uses six dimensions: content accuracy, explanatory depth, scientific reasoning, systems thinking, communication clarity and metacognitive awareness. The evaluated source configures OpenAI \texttt{gpt-4o} and Google \texttt{gemini-2.5-flash} for scoring. Model names are configuration identifiers, not immutable provider weight snapshots. The dialogue responses in the three final spoken cases identify Gemini. Stored scoring provenance includes provider/model fields and a prompt hash.

In \texttt{2d0f29bc} and later, each evaluator must return all six dimensions as JSON numbers in $[0,1]$. Missing, malformed or out-of-range values are rejected rather than silently converted or clamped. For accepted providers $P$, the resolved dimension is
\begin{equation}
s_d = \operatorname{round}_2\left(\frac{1}{|P|}\sum_{p\in P}s_{pd}\right),\qquad d\in\{1,\ldots,6\}.
\end{equation}
When both providers are available, this is their arithmetic mean. When only one valid provider remains, its scores are used with degraded-provider status; absence is not treated as zero. If neither provider supplies a complete valid set, evaluation fails. The resulting score is not ``consensus'' in the deliberative or psychometric sense.

For configured nonnegative integer weights $w_d$ summing to 100, the aggregate is
\begin{equation}
G=\operatorname{round}_2\left(\frac{\sum_d w_ds_d}{100}\right).
\end{equation}
The implementation otherwise uses the unweighted mean. A display value of $100G$ is convenient but does not convert the AI scale into a validated human grade. In the rehearsal, weights were $(30,25,20,10,10,5)$ in the dimension order above. The two-provider difference vector and a configured threshold of 0.20 were retained as review signals. Schema validity, score agreement and assessment validity are separate properties.

\subsection{Human review and candidate research safeguards}
AI evaluation moves the exercised records to a pending-review state. Educator review and publication are distinct operations. An independent-review interface begins with blank scores and suppresses AI score prefill. The later candidate requires complete integer human ratings on a 0--100 scale and retains the machine values separately. Scale labels alone do not establish equivalence; research use needs a faculty-approved mapping.

The later candidate also introduces insert-only first-pass rating receipts bound to the session, transcript hash, rubric hash and scale. Identical retries preserve the original record; conflicting submissions are refused. This protects the identity of a first-pass record in tested paths but does not supply a complete correction or adjudication history. One additional reviewer slot is not a general research-rater allocation system, and interface masking does not prove that a rater was never exposed to other information.

Research access checks bind an active consent record to the learner, study, ethics reference and exact document version/hash; withdrawal and pending-deletion states affect eligibility. The candidate builder requires explicit document metadata and configured retention days. These controls record and apply authorised study choices; software fields cannot establish that a study is ethically approved or that every operating procedure follows its approval.

A separate candidate JSON export uses explicit study scope and study-specific authorisation in addition to the administrative role. It filters consent/document and withdrawal/deletion states, rejects inconsistent participant-code mappings, includes incomplete records explicitly and separates AI evaluations, human ratings and recording metadata. It omits names, account identifiers, raw transcript and audio content, rationales and storage paths. Coded data remain potentially re-identifiable through separately held mappings and contextual knowledge. Treatment assignment and assessment phase still require a controlled external register.

\FloatBarrier
\section{Evaluation Method and Evidence Provenance}
\subsection{Engineering questions and units of analysis}
Four questions organise the analysis: can the system preserve a prerecorded response through transcription, saved state and audio retrieval; can scoring return complete inspectable dimensions and disagreement indicators; can publication and export retain the expected synthetic records; and which research safeguards have verification evidence at each release?

The units are a fixture execution, an audio artefact, an assessment record and a local test assertion. They are not interchangeable. The evaluation has four unique fixtures, two release runs, six scored executions and eight audio retrievals. Local assertion counts describe the test suite's checks, not independent trials or a population sample. No inferential statistics, learner treatment effects or reliability coefficients are estimated.

\subsection{Releases and evidence sources}
Table~\ref{tab:versions} prevents evidence from being transferred silently between revisions. The final prerecorded end-to-end rehearsal was run against a tagged production candidate; that tested image was subsequently promoted to normal public traffic. Its source snapshot is \texttt{2d0f29bc}. A later safeguard snapshot, \texttt{e2ee243}, was deployed to staging and a production candidate receiving no ordinary public traffic. The latter has local test, endpoint-smoke and limited authenticated interface evidence, not a new full end-to-end research rehearsal.

\begin{table}[htbp]\centering\small
\caption{Release-specific evidence available on 25 September 2026.}\label{tab:versions}
\begin{tabularx}{\textwidth}{p{28mm}YY}\toprule
Snapshot & Evidence used in this report & Not established by that evidence\\\midrule
Pre-update baseline & Four prerecorded inputs; transcription, dialogue, scoring and retrieval records & A controlled comparative experiment.\\
\texttt{2d0f29bc} & Same four fixtures; saved-state recovery; duplicate evaluation checks; synthetic publication and classroom export; later one-clip browser playback & Physical microphone capture; valid human grades; a research-ready export.\\
\texttt{e2ee243} candidate & Local safeguard/regression logs; staging and candidate smoke checks; limited authenticated builder inspection & Full live consent-to-export workflow; classroom capacity; all asynchronous race conditions.\\\bottomrule
\end{tabularx}
\end{table}

The report derives observations from retained request/response records, original and retrieved WAV files, synthetic classroom export reconciliation, source snapshots, deployment receipts and local test logs. Audio hashes were independently recalculated during manuscript preparation. The supplement contains an allowlisted extraction of fixture text, recognised text, scores, provider-status fields, disagreement and hashes; operational account identifiers and service configuration are not included. The complete deployment and regression bundle remains in private project storage.

\subsection{Fixture construction and execution}
The three speech inputs represent a supported explanation, a limited but broadly correct explanation and explicit misconceptions about bacterial DNA and chloroplasts. Each begins with a spoken statement that it is a synthetic technical test. A fourth input contains three seconds of silence. Appendix~\ref{app:fixtures} provides the exact text, source, question and scoring configuration so that the content test is inspectable. The scripts were machine voiced; they do not sample accents, languages, disabilities, hesitation patterns or authentic student misconceptions.

Each run used a dedicated disposable learner and educator account and four separate assessment records. Research participation was not asserted for these accounts. The harness uploaded prerecorded WAV files to the assessment pipeline, inspected transcription and dialogue, persisted and reloaded the spoken-case checkpoint, requested evaluation and retrieved retained audio. For the final three spoken cases it also retried evaluation, exercised review draft/publication, inspected learner results and reconciled the dedicated educator's classroom CSV. Those publication tests copied AI values mechanically with explicit synthetic-only notices. No independent human rating occurred.

The silence case returned empty transcription and was deliberately not submitted for evaluation. Consequently, it tests an empty transcription path and audio retention, not automatic browser rejection of a silent answer. Timing, concurrency, cost and failure-rate claims are outside the retained protocol. The two release runs use the same content but differ in system state; their score changes are not attributable solely to a specific code change or solely to model nondeterminism.

\section{Results}
\subsection{Recording integrity and exercised workflow operation}
All eight retrieved WAV files matched the SHA-256 hashes of their corresponding inputs, covering speech and silence across both runs. The three spoken cases in each run had verified saved-checkpoint recovery. In the final release, all three duplicate evaluation requests reported an already-submitted state without a repeat evaluation in the exercised path. This does not establish idempotency for every upload or transcription retry.

The final dedicated educator CSV contained exactly the three mechanically published synthetic records. It was an identified, completed-only classroom export, not a coded research export. Its row count demonstrates the scoped reconciliation exercised here, not completeness of the newer candidate research package. One post-promotion browser playback reached the end of a 32.795-second clip without a media error; a second media element in that observation was unloaded and is not counted as a successful playback.

\subsection{Scores and model disagreement}
Table~\ref{tab:scores} gives the recorded aggregates. The expected qualitative ordering occurred in both runs, but the fixtures were deliberately constructed to differ strongly in content. They are not an independently labelled benchmark and do not support an accuracy percentage.

\begin{table}[htbp]\centering\small
\caption{Aggregate outputs for the same fixtures in two release runs. Values are $100G$, not human grades.}\label{tab:scores}
\begin{tabularx}{\textwidth}{Yrrp{42mm}}\toprule
Fixture & Baseline & Final release & Final provider disagreement\\\midrule
Supported explanation & 100 & 100 & Not flagged; maximum 0.00\\
Partial explanation & 56 & 62 & Flagged; maximum 0.60\\
Misconception & 3 & 8 & Flagged; maximum 0.80\\
Silence & --- & --- & Not evaluated by the harness\\\bottomrule
\end{tabularx}
\end{table}

Both providers were reported available in all six scored executions. In the three final-release executions, both providers additionally passed the complete numeric score-schema checks. The baseline used less strict score handling, so its provider status does not establish this property. Two of the three spoken cases were flagged for disagreement in each run. The final dimension scores in Table~\ref{tab:dimensions} show why a low aggregate cannot substitute for inspection: the misconception case received zero on the four content/reasoning dimensions but nonzero clarity and metacognitive scores. These dimensions also had provider differences of 0.80. The supported short explanation received full scores on all dimensions, including metacognition, which warrants faculty calibration rather than a claim of perfect understanding.

\begin{table}[htbp]\centering\small
\caption{Resolved AI dimensions in the final release rehearsal, on a 0--1 scale.}\label{tab:dimensions}
\begin{tabularx}{\textwidth}{Yrrr}\toprule
Dimension (fixture weight) & Supported & Partial & Misconception\\\midrule
Content accuracy (30) & 1.00 & 0.78 & 0.00\\
Explanatory depth (25) & 1.00 & 0.40 & 0.00\\
Scientific reasoning (20) & 1.00 & 0.70 & 0.00\\
Systems thinking (10) & 1.00 & 0.40 & 0.00\\
Communication clarity (10) & 1.00 & 0.80 & 0.60\\
Metacognitive awareness (5) & 1.00 & 0.60 & 0.40\\\bottomrule
\end{tabularx}
\end{table}

The final partial response yields an unrounded weighted value of 0.624, displayed after rounding as 0.62. The misconception response yields 0.08. Their changes from the baseline values are release-rehearsal observations, not improvement in assessor quality or learner performance. No repeated-sampling stability experiment, model ablation, blinded human reference, agreement coefficient or subgroup analysis was conducted.

\subsection{Candidate safeguard verification}
The newer candidate's local logs record the passing checks in Table~\ref{tab:tests}. Their counts are reported separately because the suites exercise overlapping code and use different units. They must not be summed into a statistical reliability estimate. Fixtures were isolated and synthetic; these new checks made no live provider calls and used no participant data.

\begin{table}[htbp]\centering\small
\caption{Local verification recorded for the later safeguard candidate.}\label{tab:tests}
\begin{tabularx}{\textwidth}{Yrp{61mm}}\toprule
Suite & Recorded count & Scope and caution\\\midrule
Study safeguards & 66 checks & Consent/version gating, rating receipts and scoped export cases.\\
Security/retention & 61 checks & Selected authorisation and retention paths, not a complete security audit.\\
Review regression & 105 assertions & Rating validation and review-state behaviour.\\
Provider scores & 108 assertions & Complete numeric score contracts, not model accuracy.\\
Recording lifecycle & 27 tests & Recorder state/finalisation cases, not physical microphone tests.\\
Existing contracts & 4 suites & Educator workflow, submission/results, security and retrieval/speech contracts.\\\bottomrule
\end{tabularx}
\end{table}

Changed PHP files also passed syntax checks. SQLite fixtures support conclusions about those local test cases, not MySQL deployment concurrency or classroom load. Staging and the zero-traffic candidate returned successful login/assets responses and rejected unauthenticated access to the tested protected routes. An authenticated builder inspection verified explicit document fields and corrected configurable retention wording. None of these smoke checks is a substitute for a complete role-specific consent, recording, rating, withdrawal and export rehearsal.

\FloatBarrier
\section{Discussion and Limitations}\label{sec:limits}
\subsection{Operational evidence and assessment validity}
The strongest supported result is continuity across exercised technical paths: a controlled spoken input can be retained, transcribed, scored, reviewed mechanically and reconciled with its exported record. Provenance and missingness fields can help investigators discover a broken chain. They do not make the underlying score correct. Validity concerns the intended interpretation and use of scores, not merely their production \cite{kane}. Human judgement, source quality, question design, transcription error and model behaviour all influence whether the evidence supports an educational interpretation. Human oversight also depends on interaction design; evidence from AI-assisted decision-making motivates testing independent judgement rather than assuming an approval button prevents overreliance \cite{bucinca}.

The score pattern is useful as a smoke test but weak as a validation dataset. There are only three distinct speech scripts, the scripts were intentionally separated in apparent quality, and their synthetic preamble may itself influence the models. Two provider outputs are not two independent human raters. The two models can share training biases and rubric interpretations; low disagreement can accompany shared error. The uncalibrated full metacognitive score for a short supported script illustrates why aggregate ordering alone is insufficient.

The protocol did not compare the generated scores with independently assigned faculty scores, validate the six-dimensional construct structure, test accommodations or estimate learner gains. Reference ASR word-error rates were absent. No result supports claims of objectivity, bias-free assessment, instructor replacement or a validated high-stakes oral examination.

\subsection{Recovery, concurrency and operational readiness}
Reliable study collection must include interruption recovery, not only the successful path. Durable recovery of completed but unsent audio, idempotent recording ingestion/transcription, conflict-safe checkpointing and multi-tab handling remain engineering gates. In the reviewed interface, transcript recovery uses plaintext browser-local storage and failed audio remains memory-only; shared-device exposure and reload loss are unresolved concerns. Saved transcript recovery must be reconciled with cleanup, consent and retention policy. A final-write safeguard in one evaluation path does not establish complete serialisation across all capture, checkpoint, review, publication and withdrawal operations.

The study team still needs a physical-room rehearsal using the intended microphone, headphones, browser and network. It must cover quiet and noisy input, denied microphone permission, interrupted uploads, session termination, replay and two-person device turnover. Browser viewport checks are not tests on real mobile hardware and do not certify accessibility. No sustained-load or approximately 50-person concurrency experiment is reported.

The candidate's research export and first-pass rating receipts improve inspectability but leave operational requirements. A complete rater allocation and append-only correction/adjudication procedure is needed. Missing records must remain distinguishable from zero scores. Export completeness must be reconciled against a controlled enrolment and session register, including incomplete and withdrawn cases under the approved handling rules. Until these checks are satisfied, the present evidence is not participant-study launch clearance.

\subsection{Research governance and prospective evaluation}
This report analyses synthetic technical tests only. The study team reports that its planned human-study amendments have been approved; those approvals are not used here as evidence of system validation. Participant deployment must use the exact approved documents, processor arrangements, retention schedule, consent procedure and roles. No new application for approval is asserted to be necessary by this technical report, and no claim is made that human-participant data produced its results.

A future evaluation should separate two questions. First, does access to CALLIOPE formative practice improve a common, unaided learning outcome relative to an appropriate comparator? Second, how do AI scores agree with blinded faculty judgements on the same responses? A planned sample near 50 requires a justified power or precision analysis and a restrained estimand, not an assumption that it establishes broad effectiveness. Topic choice, baseline measures, assessment forms, treatment allocation and analysis rules should be fixed prospectively. Oral-project assessment and individual topic learning should not be pooled without a defensible design.

Independent raters should use a specified rubric and record first-pass judgements before reconciliation. Agreement analysis should retain the rater identity and distinguish absolute agreement from consistency; the selected intraclass correlation model and uncertainty must match the design \cite{koo}. Correlation alone is insufficient to establish interchangeability. Student learning, assessor agreement, recording reliability and participant experience are distinct outcomes. These are proposed validation activities, not completed findings.

\subsection{Reproducibility and contribution boundary}
The appendix and supplement disclose the controlled inputs and selected observations. Public availability of the live service does not reproduce a historical release. Provider model identifiers may change behaviour over time, and the private source/deployment bundle is not a public runnable release. An independent rerun would additionally require authorised software access, provider configuration, the precise runtime, audio-generation settings, deployment state and the complete harness. The retained evidence therefore supports audit of reported observations more strongly than bit-for-bit regeneration of model outputs.

The contribution is the integration and bounded evaluation of an inspectable oral assessment workflow, not proof that every component is novel. It extends a conventional chat-and-score interaction with source versioning, separate review states, recording provenance and candidate research controls. Its value for learning and consequential assessment remains an empirical question. The planned educational study should cite this technical account and report its own frozen version, methods and outcomes, rather than reusing synthetic counts as learner evidence.

\section{Conclusion}
CALLIOPE links versioned source material, oral response capture, transcription, adaptive questioning, two-provider rubric scoring and educator review. Retained synthetic rehearsals demonstrate specific recording, saved-state, scoring, duplicate-evaluation and export paths. They also expose substantive limits: provider disagreement, unvalidated score interpretation, incomplete interruption recovery and the absence of a full live research workflow evaluation for the later safeguards candidate. This report provides a technically traceable basis for further engineering and controlled study preparation. It does not establish learning gains, reliable human-equivalent grading or unconditional readiness for participant use.

\section*{Declarations}
\textbf{Authorship and contributions.} The provisional author line follows the submitting author's supplied reference report and account identity. Final author order, contributor roles and approval of public posting remain to be confirmed before submission. No collaborator is named as an author or acknowledger without confirmation.

\textbf{Funding and competing interests.} Author confirmation is pending. The report concerns software developed by the submitting author; this development role should be retained in the final disclosure. No assertion of absent financial or other interests is made in this draft.

\textbf{Data and software availability.} The public-facing system is at \url{https://calliopespeak.com}. The accompanying synthetic-observation JSON/CSV and Appendix~\ref{app:fixtures} document selected technical inputs and results. No human-participant dataset is reported or released. The full source, recordings and operational evidence are retained in private project storage; no public repository, open-source licence or independent reproducibility certification is claimed.

\textbf{AI assistance.} OpenAI Codex assisted with implementation-related technical verification, synthetic QA preparation, source/evidence inspection, literature lookup, numerical checks, LaTeX drafting and manuscript preparation. The report was prepared in an OpenAI Prism project. Generated synthetic inputs are explicitly distinguished from participant data. AI is not an author. Human author verification of all claims, references, rights and declarations is required before public submission.

\clearpage
\appendix
\section{Synthetic Fixture Specification}\label{app:fixtures}
The following text reproduces the controlled fixture inputs. These are not participant utterances. Their names describe intended content differences rather than independently validated grades. All three speech scripts include the same synthetic-test preamble, which limits ecological validity.

\subsection*{Assessment source}
\begin{quote}\small
SYNTHETIC TECHNICAL QA SOURCE, NOT COURSE MATERIAL. Bacterial cells contain DNA but lack a membrane-bound nucleus. Eukaryotic cells enclose most DNA within a nucleus and contain membrane-bound organelles. Cellular compartments support distinct local conditions; lysosomes maintain acidic conditions for breakdown reactions. Absence of a nucleus does not imply absence of DNA.
\end{quote}
\subsection*{Question and configuration}
\begin{quote}\small
Synthetic technical test: compare bacterial and eukaryotic cells, explain why cellular compartments matter, and address the claim that bacteria have no DNA.
\end{quote}
The assessment was configured as formative, with a two-minute limit and audio retention enabled. The three required concepts concern bacterial DNA without a membrane-bound nucleus, the eukaryotic nucleus and compartments, and compartment-specific conditions. Required terminology was \texttt{DNA,nucleus,compartments}. The misconception map specified that absence of a nucleus does not mean absence of DNA. Each dimension's descriptor used: ``Technical fixture only. Evaluate [dimension] using only the attached cell-comparison source. Reward correct supported reasoning; do not reward confidently incorrect claims. No real student or course grade.'' Weights appear in Table~\ref{tab:dimensions}.

\subsection*{Supported explanation}
\begin{quote}\small
This is a synthetic technical test, not a participant response. Bacterial cells contain DNA but do not have a membrane-bound nucleus. Eukaryotic cells enclose most of their DNA in a nucleus and have membrane-bound organelles. Compartments allow different processes to operate under suitable conditions. For example, lysosomes maintain an acidic environment for breakdown reactions. Saying that bacteria have no DNA confuses the absence of a nucleus with the absence of genetic material.
\end{quote}
\subsection*{Partial explanation}
\begin{quote}\small
This is a synthetic technical test, not a participant response. Eukaryotic cells have a nucleus. Bacterial cells do not have a membrane-bound nucleus. I think separate parts do different jobs, but I cannot yet explain why that helps the cell.
\end{quote}
\subsection*{Misconception}
\begin{quote}\small
This is a synthetic technical test, not a participant response. I believe that bacteria have no DNA because they do not have a nucleus. I also think every eukaryotic cell has chloroplasts. These are my current claims, and I cannot provide supporting evidence.
\end{quote}
The silent control contains three seconds of silence and no input text. It was retained and retrieved but not submitted for scoring. The machine-readable supplement preserves recognised text separately from these inputs, with null scores for ungraded cases. It contains four fixtures per run; repeated content is not an independent sample.

\section{Evidence Scope and Release Checklist}
\begin{table}[H]\centering\small
\caption{Claim-to-evidence map for interpreting this report.}\label{tab:claims}
\begin{tabularx}{\textwidth}{YY}\toprule
Supported narrow claim & Additional evidence needed for a stronger claim\\\midrule
Eight retrieved synthetic files match their inputs & Real capture, interrupted ingestion, storage lifecycle and multi-device recovery tests.\\
Three final spoken fixtures have complete two-provider scores & Blinded faculty reference, repeated model runs, construct and subgroup validity.\\
Three synthetic classroom rows reconcile & Full coded research export reconciliation including consent, missingness and withdrawal.\\
Candidate local checks pass & Full deployed role-specific workflow and production-database concurrency tests.\\
Responsive visual fixture is retained & Physical devices, assistive technology, user accessibility and usability testing.\\
First-pass receipts bind reviewed evidence in local tests & Complete rater allocation, correction/adjudication and observed blinding procedures.\\\bottomrule
\end{tabularx}
\end{table}
Before public posting, the author must verify the final author list and permissions, funding and interest disclosures, manuscript content, reference metadata and the chosen distribution licence. Before participant launch, the team must separately complete the engineering, approved-document, course-material, rater, room/device and export gates. Public preprint posting is not a substitute for either review.
\end{document}